\documentclass[pre,english,aps,a4paper,twocolumn,showpacs,floatfix]{revtex4}
\usepackage[T1]{fontenc}
\usepackage{amsmath}
\usepackage{amsfonts}
\usepackage[latin1]{inputenc}
\usepackage[usenames]{color}
\usepackage{graphicx}
\usepackage{epsfig}
\begin{document}
\title{Surface and smectic layering transitions in binary mixtures of parallel 
hard rods}
\author{Daniel de las Heras} \email{daniel.delasheras@uam.es}
\affiliation{Departamento de F\'{\i}sica Te\'orica de la Materia Condensada,
Universidad Aut\'onoma de Madrid, E-28049 Madrid, Spain}
\author{Yuri Mart\'{\i}nez-Rat\'on}\email{yuri@math.uc3m.es}
\affiliation{Grupo Interdisciplinar de Sistemas Complejos (GISC),
Departamento de Matem\'{a}ticas,Escuela Polit\'{e}cnica Superior,
Universidad Carlos III de Madrid, Avenida de la Universidad 30, E--28911, Legan\'{e}s, Madrid, Spain}
\author{Enrique Velasco}\email{enrique.velasco@uam.es}
\affiliation{Departamento de F\'{\i}sica Te\'orica de la Materia Condensada
and Instituto de Ciencia de Materiales Nicol\'as Cabrera,
Universidad Aut\'onoma de Madrid, E-28049 Madrid, Spain}
\date{\today}
\begin{abstract}
The surface phase behavior of binary mixtures of colloidal hard rods 
in contact with a solid substrate (hard wall) is studied, with special
emphasis on the region of the phase diagram that includes the
smectic A phase. The colloidal rods are modelled as hard cylinders
of the same diameter and different lengths, in the approximation of perfect 
alignment. A fundamental--measure density functional is used to obtain 
equilibrium density profiles and thermodynamic properties such as
surface tensions and adsorption coefficients. The bulk phase diagram
exhibits nematic-smectic and smectic-smectic demixing, with smectic
phases having different compositions; in some cases they are
microfractionated. The calculated surface phase diagram of the 
wall-nematic interface shows a very rich phase behavior, including
layering transitions and complete wetting at high pressures, whereby
an infinitely thick smectic film grows at the wall via an infinite 
sequence of stepwise first--order layering transitions. For lower pressures 
complete wetting also obtains, but here the smectic film grows in a continuous
fashion. Finally, at very low pressures, the wall-nematic interface exhibits 
critical adsorption by the smectic phase, due to the second-order character 
of the bulk nematic-smectic transition.  
\end{abstract}


\maketitle

\definecolor{Red}{rgb}{1.00,0.00,0.00} 

\section{Introduction}

The wetting behavior of molecular smectic (S) liquid crystals in contact 
with a solid substrate \cite{Ocko1,Lucht,Moses,Jin,Lau} or at their vapor-liquid 
crystal interfaces \cite{Ocko2,Pershan,Lucht1,Lucht2,Fukuto} has been an active 
research area since the 80's. Partial or complete wetting behaviors of the S phase 
have been found when the isotropic (I) or nematic (N) phases are stable at bulk. 
Some liquid crystals exhibit a sequence of first-order layering transitions   
on decreasing the temperature slightly above the IS or NS bulk transition 
temperatures \cite{Ocko1,Lucht,Moses,Jin,Lau,Ocko2,Pershan,Lucht1,Lucht2,Fukuto},
indicating that the partial and complete wetting r\'egimes can be mediated by a
finite (or infinite) sequence of stepwise layer-adsorption transitions.

Usually the wetting behavior depends on specific interactions between the 
surface and the liquid crystal molecules, and molecular characteristics such as 
dipoles or length of alkyl chains. Depending on the strength of the interactions, 
solid substrates inducing orientational ordering of molecules may favor partial or 
complete wetting of the substrate by the N or S phases when the I phase is stable at 
bulk, while those substrates promoting orientational disorder favor partial wetting 
behavior. 

Freely-suspended smectic films, consisting of a few smectic layers surrounded
by vapor (V), are formed by some liquid crystals and constitute 
another example of phase transitions induced by the presence of a surface. 
These films exhibit the so-called thinning transitions whereby 
the film thickness decreases stepwise as one or several layers (depending
on the film heating rate) melt \cite{Stoebe,Jin2,Demikhov}.  

As usual in statistical mechanics, lattice models were the first to be
applied to the study of the surface phase behavior of smectic liquid crystals. 
For example, a version of the Lebwohl-Lasher model, extended to include the smectic 
phase, was used to study the systematics of layering phenomena  
\cite{Pawlowska}. 

Density-functional theory (DFT) has been also
successfully applied to the study of the surface phase behavior in
liquid crystals adsorbed on solid substrates.
The extension of the MacMillan theory to non-uniform phases with the inclusion 
of surface interaction potentials accounted for layering 
and thinning transitions \cite{Selinger,Mirantsev}. However, DFT models
that incorporate repulsive interactions (reflecting molecular volume and shape)
using either the Local- (LDA) \cite{Margarida} of Weighted-Density Approximation
(WDA) \cite{Mederos}, plus anisotropic attractive interactions via a mean-field 
approximation, turned out to be more realistic models for the calculation of 
surface phase diagrams. This is due to the fact that (i) the liquid
crystal bulk phase behavior (e.g. values of coexistence densities and 
orientational order parameters) is better calculated from DFT,  
and (ii) interfacial properties, such as the width of the interface or the 
oscillatory behavior of the density profiles, are much better accounted 
for, due to the proper inclusion of pair correlations between particles.

For example, the wetting behavior of a smectic film in contact with an attractive 
wall has been successfully studied in Ref. \cite{Somoza}, where the authors found 
complete or partial wetting by smectic depending on the strength of the external potential. 
A infinite (complete wetting) or finite (partial wetting) sequence of
layering transitions was found, some of them ending in a prewetting line. 
Layering transitions at the V-I interface near the V-I-S triple
point, and thinning transitions in freely-suspended smectic films,
have been successfully studied using similar versions of DFT based on WDA 
and perturbation theory \cite{M-R0}. 

{Finally, recent theoretical works have applied related models for hard rods
in contact with a wall and/or confined between two walls. These studies were
based on different approximations: Onsager with restricted orientations
\cite{Evans1,Evans2}, Onsager with Parsons-Lee rescaling and free
orientations \cite{Dani0,Dani0a} and also a WDA functional approximation
\cite{Dani1,Dani2}. The surface phase diagram of a fluid of hard 
spherocylinders in contact with a single wall promoting different surface 
anchoring was analysed in Ref. \cite{Dani0}. In Refs. \cite{Dani1,Dani2}
the surface phase diagram obtained for the confined fluid
includes capillary nematization and 
smectization of the fluid, and a sequence of layering transitions of the 
confined smectic as the width of the slit pore is changed.}
 
Practically all the experimental work on the wetting behavior 
of liquid crystals has been focused on one-component systems, 
the extension to mixtures being a pending issue. Adsorption phenomena in
liquid crystal mixtures have a fundamental interest since bulk
demixing transitions between two phases, at least one of them being smectic, 
would add much more complexity to the surface phase behavior. 
A recent theoretical work, based on Onsager theory, has analysed the 
phase behavior of the I-N interface of binary mixtures of hard spherocylinders 
\cite{Shundyak}. Also, the substrate-isotropic interface of a mixture of hard 
parallelepipeds has been studied within the Zwanzig approximation \cite{Harnau}. However, 
it would be interesting to extend these studies to the high-pressure r\'egime, 
where the smectic phase is stable. 

One of the aims of the present work is to elucidate the role of the smectic phase 
in the interfacial phase behavior of binary mixtures. Recent theoretical    
models of mixtures of colloids (spherical or rod-like) and polymers 
\cite{Brader,Evans,Roth,Bryk} (based on the model proposed 
in Ref. \cite{Bolhuis} or on the recent fundamental-measure 
functional for hard-sphere/hard-needle mixtures \cite{Schmidt})
have shown that the entropic character of particle interactions, together
with the coupling between species generated by the external surface potential,
results in a rich phase behavior. For high polymer fugacities, partial wetting 
of the interface between the substrate and the fluid poor in colloidal
particles by the fluid rich in colloidal particles was obtained.
In the partial wetting r\'egime, a sequence of up to four
layering transitions was found. At lower fugacities complete wetting is reached 
via a first-order wetting transition (located below the critical point). These  
results were confirmed by Monte Carlo (MC) simulations \cite{Dijkstra}.

Colloidal rod-like fluids and their mixtures are paradigmatic systems 
exhibiting liquid-crystal textures similar to those of molecular fluids,
but the interaction between their components have an entropic origin
due to short-ranged repulsive forces. Intense experimental work on pure
and mixed suspensions has been done in the last two decades, demonstrating
this analogy \cite{Lekker}. Also, recent work has shown the 
importance of smectic layering in the kinetics of the NS phase transition in 
colloidal rods \cite{Dogic}, confirming the analogy between molecular and 
colloidal fluids as regards the surface-enhanced smectic ordering near a bulk 
phase transition. 

The aim of the present article is to theoretically analyse the surface 
phase diagram of a colloidal binary mixture of rods with the same breath 
but different lengths $L_i$ ($i=1,2$); in the following we use a length ratio 
$s=L_2/L_1=3$, with short species being labelled as 1 and the long species as 2. 
Particles interact through a hard repulsive potential and are 
constrained to be perfectly aligned along a nematic director, with their main 
axes perpendicular to a hard wall (W), thus simulating perfect homeotropic anchoring. 
This restriction, which considerably simplifies the model, is valid as long as 
one is only interested in the surface phase behavior of particles exhibiting a high  
degree of orientational order. The study concerns the
wetting properties of these mixtures when a smectic film partially or  
completely wets the WN interface. Our theoretical tool is based on density-functional 
theory, more specifically on a recently proposed Fundamental-Measure Functional (FMF) 
for binary mixtures of parallel hard cylinders \cite{M-R1}. 

{The impact of restricted orientations was analysed by 
Shundyak and van Roij in the context of the Onsager theory \cite{SvR}, using the 
Zwanzig model (discrete orientations). It
was found to lead to spurious nematic phases with very high orientational
order. Smectic phases were not analysed by Shundyak and van Roij but spurious
smectic phases might well exist in Onsager theory. The parallel particle 
approximation implicit in the FMF approach is not expected to lead to any such
anomalous phases, since the FMF theory 
contains a much better treatment of correlations, hence 
of the ordered phases.}

As we will see later, the surface phase diagram of the model
exhibits three different wetting behaviors depending on 
the value of pressure: (i) At high pressure $p$ we find complete wetting by smectic
via an infinite sequence of layering transitions as the nematic binodal of the
bulk NS transition is approached. These layering transitions end in 
corresponding surface critical points characterised by values of critical
pressure $p_c^{(n)}$, $n=1,2,...$. (ii) For sufficiently low pressure such that 
$p<p_c^{(n)}$, $\forall n$, wetting by the smectic film becomes continuous, with 
adsorption coefficients diverging 
logarithmically. And (iii) for pressures below the tricritical point, 
where the bulk NS transition changes from first to second order, 
we find critical adsorption by smectic. In this case a modified adsorption 
coefficient diverges logarithmically on approaching the second-order bulk 
NS transition. This divergence is a direct consequence of the
NS bulk correlation length diverging at the transition. 

The article is organized as follows. Sec. \ref{theory} is devoted to the 
presentation of the theoretical model and the numerical details 
relevant for the calculation of the bulk (Sec. \ref{bulk0}) 
and surface phase diagrams (Sec. \ref{w-f}). 
In Sec. \ref{results} we present the results obtained from 
numerical functional minimization. This section is 
divided into Sec. \ref{bulk}, where the phase behavior of this particular mixture
is described, Sec. \ref{layering}, which contains a description of the
layering transitions, and Sec. \ref{wetting}, devoted to the study of the 
wetting behavior. Finally some conclusions are drawn in Sec. 
\ref{conclusions}. Two appendices are included which contain
mathematical details on the bifurcation analysis (Appendix \ref{bifurca}) and 
the derivation of the interfacial Gibbs-Duhem relation with 
composition and pressure as independent variables (Appendix \ref{G-D}).

\section{Theoretical Model}
\label{theory}

Our particle model consists of a binary mixture of parallel hard cylinders, with both  
species having the same diameter, $D_1=D_2=D$, chosen so as to set the ratio
of transverse particle area and cylinder length squared of the short species 
to unity, i.e. $\pi D^2/4L_1^2=1$. This implies a particle aspect ratio of
$L_1/D_1=0.89$. Since we choose a length ratio $s\equiv L_2/L_1=3$, the aspect
ratio of the other particle is $L_2/D_2=2.66$. 
As density-functional theory and simulations show, a one-component fluid
of parallel hard cylinders presents a phase sequence nematic-smectic-crystal,
which is independent of the aspect ratio. The smectic phase of freely rotating hard 
spherocylinders is known to begin for aspect ratios $\agt 4.1$, and we should expect 
a similar behavior for freely rotating hard cylinders. Since the phase behavior of 
a binary mixture of parallel particles with identical diameters but different lengths
depends only on the ratio $L_2/L_1$, our model might describe a freely rotating 
binary mixture of cylinders with aspect ratios $L_1/D_1>4.1$ and $L_2/L_1=3$,
both of which would have a smectic phase. Therefore, our choice guarantees that, in the 
one-component limits, the mixture would possess stable smectic phases at high enough 
pressure in the freely-rotating case.

\subsection{Bulk smectic phase}
\label{bulk0}
A fundamental-measure density-functional theory for binary mixtures, in the version 
proposed in \cite{M-R1} and tested against MC simulations in \cite{M-R2}, 
will be used in all calculations.
We will consider a mixture which presents a non-uniform structure along
the $z$ direction. The excess free-energy density reads 
\begin{eqnarray}
\Phi L_1^3=n\left\{-\ln\left(1-\eta\right)
+\frac{3\eta}{1-\eta}+\frac{\eta^2}{\left(1-\eta\right)^2}
\right\},
\label{fmt}
\end{eqnarray}
where we drop the $z$-dependence for the sake of convenience and have defined 
the weighted densities
\begin{eqnarray}
n(z)&=&\frac{1}{2}\sum_{i}\left[\rho_i^*(z-\kappa_i/2)+\rho_i^*(z+\kappa_i/2)\right], 
\label{n0}\\
\eta(z)&=&\sum_{i}\int_{z-\kappa_i/2}^{z+\kappa_i/2}\rho_i^*(z')dz',
\label{n3}
\end{eqnarray}
with $\eta(z)$ the local packing fraction of the mixture. Index $i$ in all sums is
assumed to run for $i=1,2$. We have defined the dimensionless densities 
$\rho^*_i(z)=\rho_i(z)L_1^3$, and $z$ coordinates are also in units of $L_1$.
The $\kappa_i$ parameter is the particle length of species $i$ in the same
units. Our choice for $L_i$ gives $\kappa_1=1$ and $\kappa_2=s=3$. 
The free-energy functional per unit area can be  
calculated as $\beta {\cal F}/A=\int dz \left[\Phi_{\rm{id}}(z)+\Phi(z)\right]$,
with $\beta^{-1}=k_BT$ the inverse thermal energy and 
\begin{eqnarray}
\Phi_{\rm{id}}(z)=\sum_i \rho_i(z)\left[\ln {\cal V}_i\rho_i(z)-1\right],
\end{eqnarray}
the ideal part of the free-energy density, where ${\cal V}_i$ is the 
thermal volume of species $i$. Now we specify for the smectic phase, which is
the lowest symmetry phase considered in this work and has the property
$\rho_i(z+kd)=\rho_i(z)$ (with $d$ the smectic period and $k\in\mathbb{Z}$). The
pressure of the mixture can be calculated as
\begin{eqnarray}
&&\beta pL_1^3=d^{-1}\int_0^d\left\{
\frac{n(z)}{1-\eta(z)}+\frac{3n(z)\eta(z)}{\left[1-\eta(z)\right]^2}
\right.\nonumber\\
&&\hspace{3.8cm}\left.+\frac{2n(z)\eta(z)^2}{\left[1-\eta(z)\right]^3}\right\}dz.
\label{pp}
\end{eqnarray} 
During the numerical minimization we have used the following constraints: 
(i) the value of the pressure $p$ is fixed, and (ii) the composition of the
mixture, $x\equiv x_1=\rho_1/\rho$, is also set in advance. Here $\rho=\rho_1+\rho_2$ is
the total mean density (calculated from the constant-pressure constraint), 
while $\rho_i=d^{-1}\int_0^d \rho_i(z)dz$ is the mean 
density of the $i$-th species. The Gibbs free-energy functional 
per particle, defined as
\begin{eqnarray}
\beta g[\rho_1,\rho_2]=\rho^{-1}\left\{
d^{-1}\int_0^d\left[\Phi_{\rm{id}}(z)+\Phi(z)\right]dz +
\beta p\right\},
\end{eqnarray} 
has been minimized with respect to the densities $\rho_i(z)$. We do this
numerically by first discretising the densities, defining a grid with points
$z_k=z_0+k\Delta$ ($k=0,...,N$), and then minimising the function
$g(\boldsymbol{\rho}_1,\boldsymbol{\rho}_2)$ with respect to the components 
of the vectors 
$\boldsymbol{\rho}_i=\left[\rho_i(z_0),\cdots,\rho_i(z_N)\right]$, and 
also with respect to $d$, using a conjugate-gradient algorithm. $N$ is
the number of grid intervals. The width of the intervals
was taken to be $\Delta/L_1=0.01$, and $N\Delta=md$, where $m$ is the number 
of smectic periods within the minimization box. Varying $x$ between 0 and 1 and using 
the common-tangent construction for the function $\beta g(x)$, we have
calculated the coexistence values for $x$ and $\rho$. Repeating the above 
procedure for different pressures, we obtained the demixing binodals. 

When the NS transition is of the second order, one can use a bifurcation 
analysis to find the total packing fraction $\eta=\sum_i\rho_i^* \kappa_i$ 
and the smectic period $d$ at bifurcation {(the local and total packing
fractions are equal in the nematic phase. In the smectic
phase the average of the local fraction $\eta(z)$ over one period gives the
total packing fraction $\eta$)}. Also, with
the aim to check the relative stability of the S with respect 
to the columnar (C) phase, we extended the bifurcation analysis 
to include the columnar symmetry. For this purpose we need 
to solve the following set of equations:
\begin{eqnarray}
{\cal H}({\bf q},\eta)=0, \quad \boldsymbol{\nabla} {\cal H}({\bf q},\eta)=0,
\label{set}
\end{eqnarray}
where the wave vectors ${\bf q}=({\bf 0},q)$ and ${\bf q}=({\bf q}_{\perp},0)$ 
are appropriate for the S and C symmetries, respectively. These equations have
to be solved for the absolute minimum of ${\cal H}({\bf q},\eta)\equiv
\text{det}[H({\bf q},\eta)]$ as a function of ${\bf q}$, with
$H({\bf q},\eta)$ a $2\times 2$ matrix defined by the elements
\begin{eqnarray}
H({\bf q},\eta)=
\left(
\begin{matrix}
1-\rho_1\hat{c}_{11}({\bf q},\eta) & -\rho_1\hat{c}_{12}({\bf q},\eta)\\\\
-\rho_2\hat{c}_{12}({\bf q},\eta) & 1-\rho_2\hat{c}_{22}({\bf q},\eta)
\end{matrix}
\right),
\end{eqnarray} 
with $\hat{c}_{ij}({\bf q},\eta)$ the Fourier transforms of the direct 
correlation functions calculated from the second functional derivatives 
of the free energy functional $\beta {\cal F}[\{\rho_i\}]$ with respect to  
$\rho_i({\bf r})$ and $\rho_j({\bf r}')$. Expressions for 
these functions and explicit results for the NS and NC spinodals can be found 
in Appendix \ref{bifurca}.

\subsection{Wall-fluid interface}
\label{w-f}

The aim is to calculate the equilibrium density profiles of the two species 
in the presence of a hard wall. The wall is located at $z=0$
{and the long axes of cylinders are perpendicular to the
wall. Thus, perfect homeotropic alignment of the nematic director is assumed. 
This model may apply to experimental systems where homeotropic anchoring is 
forced by surface treatment \cite{Vaughn,Yi,Zhao,Beica} or by the application of an external field
(see \cite{Chaikin} for an example on colloidal discs).

The values of the chemical potentials of the two components, $\mu_i$, will be fixed,
which means that the conditions of the bulk fluid, far from the wall,
will be specified in advance
and maintained fixed during the minimisation. We minimise 
the grand potential functional per unit area,
\begin{eqnarray}
\frac{\Omega[\{\rho_i\}]}{A}=\frac{{\cal F}[\{\rho_i\}]}{A}+
\sum_i\int\left[v_i(z)-\mu_i\right]\rho_i(z) dz,
\end{eqnarray} 
with respect to the density profiles $\rho_i(z)$. The external potentials are
defined by 
\begin{eqnarray}
\beta v_i(z)=\left\{
\begin{matrix}
\infty, & z\leq L_i/2,\\\\
0, & z>L_i/2,
\end{matrix}
\right.\hspace{0.4cm}i=1,2.
\end{eqnarray}
To numerically implement the minimization we proceed
by first choosing values for the pressure $p$ and the composition of the mixture at
bulk, $x$, and from here calculating the chemical potentials $\mu_i$ and the 
dimensionless total density $\rho^*=\rho L_1^3$ at an infinite distance from the wall, 
using the following expressions, which apply to the bulk nematic phase:
\begin{eqnarray}
\beta pL_1^3&=&\frac{\rho^*(1+\eta)}{(1-\eta)^3},
\label{press}
\end{eqnarray}
and
\begin{eqnarray}
\beta \mu_i&=&\ln x_i+\ln\left(\frac{\rho^*}{1-\eta}\right)
+\frac{\eta(3-2\eta)}{(1-\eta)^2}\nonumber\\\nonumber\\
&+&\frac{\rho^*(4-3\eta+\eta^2)}{(1-\eta)^3}\kappa_i,
\quad i=1,2.\label{chepo}
\end{eqnarray}
The implicit Eqn. (\ref{press}) has to be solved iteratively to obtain $\rho^*$.
In the minimisation the usual boundary conditions at a large distance $H$ from
the wall, $\rho_i(H)=x_i \rho$, have to be imposed. $H$, the width
of the minimization box, is chosen in such a way as to guarantee 
that the structure of the WN interface can be accommodated within the box
and at the same time the boundary conditions are satisfied. 
Finally, the surface tension of the interface is calculated as 
$\gamma_{\hbox{\tiny WN}}=\Omega[\{\rho_i^{(e)}\}]/A+pH$,
with $\rho_i^{(e)}$ the equilibrium density profiles. 

One of our aims is to obtain the wetting behaviour of the mixture when 
nematic conditions are fixed at bulk and the NS demixing transition is
approached. This means that we need to calculate the surface tension of the 
WS interface for values of the chemical potentials $\mu_i$ corresponding to 
NS coexistence. Therefore $\mu_i$ can be calculated from Eqn. (\ref{chepo}). 
However, if the bulk phase is a smectic, and consequently the density profiles 
are not uniform in bulk, the boundary conditions depend on the particular 
value of $H$ chosen, which considerably complicates the numerical minimization. 
To avoid this problem, we choose to define a symmetric box by using the following 
external potentials: 
\begin{eqnarray}
\beta v_i(z)=
\left\{
\begin{matrix}
\infty, & z\leq L_i/2,\\\\
0, & L_i/2<z<H-L_i/2,\\\\
\infty, & z\geq H-L_i/2.
\end{matrix}
\right.
\end{eqnarray}
We minimize $\Omega[\{\rho_i\}]$ with respect to $\rho_i(z)$ by choosing $H$ large 
enough to accommodate two WS interfaces. However, due to long-ranged 
commensuration effects generated by the confinement of a layered phase with
period $d$ in a slit of width $H$, the minimized grand potential 
exhibits an oscillatory behavior as a function of $H$, with an asymptotically 
decaying amplitude. To overcome this problem, we defined the curve obtained from 
the local minima of $\Omega[\{\rho_i^{(e)}\}]/A$ and extrapolated to $H\to\infty$
to obtain the value of $2\gamma_{\hbox{\tiny WS}}$ 
(i.e. two times the surface tension of the WS interface). 

To find the surface tension of the NS interface we followed a similar 
procedure: we defined a box of width $H$ with boundary conditions 
$\rho_i(0)=\rho_i(H)=\rho_i^{(\hbox{\tiny N})}$ (the densities of the bulk nematic phase 
coexisting with smectic) at both ends of the box. Choosing an initial guess for 
$\rho_i(z)$, $0<z<H$ (close to the profiles of the coexisting bulk smectic phase), 
we minimized the grand potential to obtain $2\gamma_{\hbox{\tiny NS}}$ (i.e. two times the 
surface tension of the NS interface). Again the value of $H$ has to be large enough 
to accommodate two NS interfaces. Having the surface tensions of all the three 
different interfaces, one can study the wetting behavior of the system, which is
discussed in Sec. \ref{wetting}. 

Adsorption coefficients will also be used as a convenient measure of the wetting 
and adsorption properties of the WN interface. The adsorption coefficients of both
species are defined as 
\begin{eqnarray}
\Gamma_i=\int_0^H\left[\rho_i(z)-\rho_i^{(\hbox{\tiny N})}
\right]dz,\hspace{0.4cm}i=1,2.
\label{Ads}
\end{eqnarray}
In Appendix \ref{G-D} a derivation is presented of the interfacial Gibbs-Duhem 
relation expressed in terms of the independent variables $x$ and $p$. 
Using this equation, a relation between the derivative of 
$\gamma_{\hbox{\tiny WN}}$ with respect to the composition 
variable $x$ and the adsorption coefficients can be obtained:
\begin{eqnarray}
\beta\frac{d\gamma_{\hbox{\tiny WN}}}{dx}={\cal U}(x,p)\left(\frac{\Gamma_2}{1-x}
-\frac{\Gamma_1}{x}\right),
\label{relation}
\end{eqnarray}
where ${\cal U}(x,p)$, a function of bulk composition and pressure, is always 
positive if the binary mixture is stable against NN demixing. This relation has
been tested (see Appendix \ref{G-D}) to check for consistency of our numerical 
minimization procedure. Also, the sum rule relating the bulk pressure with the
densities at the wall (contact theorem), $\beta p=\rho_1(L_1/2)+\rho_2(L_2/2)$, which
is automatically satisfied by the functional, provides another
check for numerical accuracy. For example, for a mixture with bulk pressure 
$\beta pL_1^3=1$ and composition $x=0.82$, we obtain 
$(\rho_1(L_1/2)+\rho_2(L_2/2))L_1^3=0.994$, $0.997$ and $0.999$ for values of the 
discretisation interval along the $z$ axis of $\Delta L_1^{-1}=0.0100$, $0.0050$
and $0.0025$, respectively (obviously, in the limit where $\Delta z\to 0$, the sum 
rule becomes exact).

\section{Results}
\label{results}
This section is devoted to the presentation of the results obtained from 
our theoretical model. It is divided into three different sections.  
In Sec. \ref{bulk} we present and describe the main features of 
the bulk phase diagram. Secs. \ref{layering} and \ref{wetting} are 
devoted to the layering transitions and to the wetting behavior, respectively.  

\subsection{Bulk phase diagram}
\label{bulk}

The bulk phase diagram of the binary mixture, shown in Fig. \ref{fig1},
has been calculated using bifurcation analysis and density-functional minimization, 
as described in Sec. \ref{bulk0}. Two NS spinodals (dashed curves in
Fig. \ref{fig1}), calculated from the bifurcation analysis, depart from the 
one-component limits $x=0$ and $x=1$ (where, as defined above, $x$ is
the composition of the mixture as given by the fraction of short particles). 
The values of pressure in both
spinodals increase as the composition of the mixture becomes farther from these
limits, indicating that the two species cannot be easily accommodated into a smectic 
arrangement. The spinodal lines end in a tricritical point (filled circle) and a 
critical end point (filled square), respectively. 
Functional minimization indicates that the Gibbs free energy 
of the smectic phase is always a convex function of composition $x$ in 
the neighbourhood of (and above) these lines, which proves that the NS 
transition is of second order, with 
the smectic order parameter increasing from zero at the bifurcation. For pressures 
above the tricritical point but below the critical end point,
the mixture segregates into a smectic phase rich in the 
long species and a nematic phase rich in the short species.

Two different smectic phases occur in the region of smectic stability.
These phases are distinguished by the relative location of the density peaks of the 
two species. In the smectic phase labelled as S$_1$ the profiles are in phase,
with density peaks of the two species 
located at the same positions, which define the location of 
the smectic layers. In the phase called S$_2$ the density profiles are out of
phase, forming alternating smectic layers: this phase exhibits 
microfractionation \cite{Koda,Giorgio1,Giorgio2}.
The smectic mixtures with a higher fraction of species 1 (the short component) 
are denoted with a prime in Fig. \ref{fig1}. Examples of density profiles 
corresponding to these two smectic phases are shown in Figs. \ref{fig2}(a)-(d). 
Several regions of smectic coexistence exist in the mixture: S$_1$-S$_2$ in a
narrow pressure interval between a critical and a triple point,
S$_1^{\prime}$-S$_2^{\prime}$ in a corresponding interval between critical and
triple points, and S$_1$-S$_1^{\prime}$ and S$_1$-S$_2^{\prime}$ coexistences
at high pressure (triple points have been indicated by horizontal dashed lines
in the figure).

\begin{figure}
\epsfig{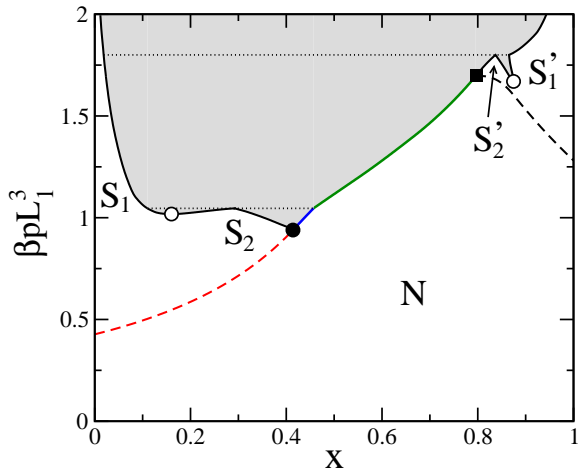}
\caption{(Colour online). 
Bulk phase diagram of the binary mixture of parallel hard cylinders 
in the reduced  
pressure $\beta pL_1^3$--composition $x$ plane (with $x=x_1$, the fraction
of short particles). Dashed curves represent second-order NS  
transitions, while solid lines are the binodals of the NS or SS demixing 
transitions. The shaded area is the region of instability. Stability regions
of nematic and different smectic phases are denoted by letters (see text).
Colour lines indicate the conjectured wetting behaviour along the
NS lines: critical adsorption
(red), complete wetting by S$_2$ smectic phase without layering transitions (blue),
and complete wetting by S$_1$ phase mediated by layering transitions (green). 
{Circles: critical points. Square: critical end-point. Triangle: tricritical point.}} 
\label{fig1}
\end{figure}

{We now comment on the possible stability of the columnar
phase. A complete calculation of the stability of the columnar phase 
by free-energy minimisation is, at present, a highly difficult task. 
The difficulties stem from the computation of two-particle weighted 
densities \cite{M-R1}. Therefore, we have implemented a bifurcation
analysis, which gives the conditions under which the nematic phase
becomes unstable with respect to columnar-like fluctuations.} 
As shown in Appendix \ref{bifurca}, the NC spinodal, signalling the instability
of the nematic phase against columnar-like fluctuations, is always 
located above the NS spinodal for all values of composition. This is an 
indication that at least part of the phases depicted in the phase diagram 
of Fig. \ref{fig1} could be stable, {and that the surface 
behaviour to be described below could represent the real surface behaviour of 
the model. However, one has to be cautious, since a first-order 
nematic-columnar and/or smectic-columnar phase transition could be 
greatly displaced with respect to the spinodal lines. 
Experimentally, rod-like colloidal particles always have some degree of
polydispersity. Diameter polydispersity would tend to destabilise the
columnar phase against the smectic phase, while breadth polydispersity
would have the opposite effect \cite{Poly}. The final balance may 
depend on several effects in a delicate manner.
Therefore, one has to be cautious until the following aspects are considered:
(i) particle polydispersity in length and breadth, and (ii) full 
minimization of the density 
functional with respect to density profiles having columnar symmetry. 
We do not pursue this analysis here, which is left for future work.}

\begin{figure}
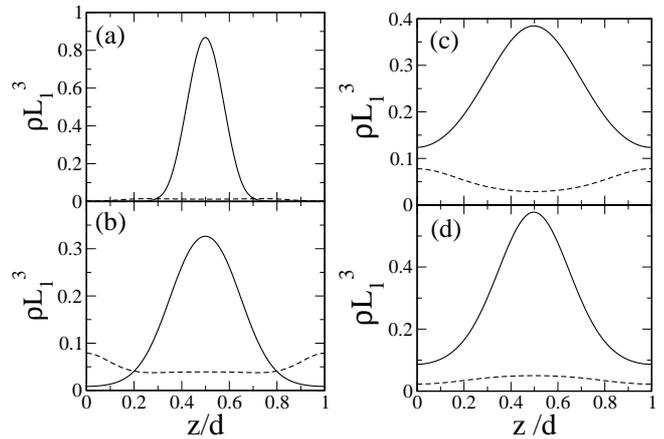

\epsfig{file=Fig2a.eps,width=1.65in}
\epsfig{file=Fig2b.eps,width=1.65in}
\caption{Density profiles of (a) S$_1$, (b) S$_2$, (c) S$_2^{\prime}$ (c), and (d)
S$_1^{\prime}$ phases in one smectic period. Values of smectic period
are: (a) $d/L_1=3.495$, (b) 3.835, (c) 1.433, and (d)
1.237. In all figures solid and dashed curves correspond to species 2 and 1, 
respectively. Values of reduced pressure and composition, $(\beta pL_1^3,x)$, are:
(a) $(1.20,0.06)$, (b) $(1.04,0.28)$, (c) $(1.75,0.83)$, and (d) $(1.75,0.88)$.} 
\label{fig2}
\end{figure}

\subsection{Layering transitions}
\label{layering}
In this section, which constitutes the cornerstone of the present work, we 
present a detailed study of the layering transitions in the mixture. The stable
bulk phase (in the region infinitely away from the wall) will be chosen to 
be a nematic phase, characterised by particular values of pressure and composition.
We first consider the case where the pressure is given a value $\beta pL_1^3=1.30$ 
and the bulk composition $x$ is decreased from a high value close to unity. 
As the nematic branch of the NS$_1$ binodal is approached, a sequence of layering 
transitions is found. At each of these transitions a new smectic layer, mostly 
composed of particles
of the long species, appears through a first-order (interfacial) phase transition. 
This is illustrated in Fig. \ref{fig3}, where four equilibrium WN interfaces 
containing 0, 1, 2 and 3 smectic layers composed (essentially) of particles of
species 2 are shown. These structures will be denoted by WN$_i$, with $i$ the number 
of adsorbed layers. 

The structure of the WN$_0$ interface is interesting. Right at the wall
there is a mixture of highly localised long and short particles with 
similar densities. For increasing distance from the wall
the density structure becomes much weaker
[see Fig. \ref{fig3}(a)]. At the first (WN$_0$-WN$_1$) layering transition,
the wall becomes fully covered by long particles and a single very high density 
peak appears [Fig. \ref{fig3}(b)]. On further decreasing $x$, the system exhibits 
a sequence of phase transitions, WN$_i$-WN$_{i+1}$,
each involving the addition of a further highly
localised peak of the long particles. At $x=x_{\rm{coex}}=0.6115$ 
(the composition of 
the bulk nematic phase coexisting with the S$_1$ phase at bulk), the wall is
completely wet by the S$_1$ phase, which means that a macroscopically thick 
smectic film (consisting of an essentially infinite number of smectic layers) is 
interposed between the wall and the nematic phase. We have found up to 12 
layering
transitions as $x\to x_{\rm{coex}}$ with $x>x_{\rm{coex}}$.
Access to higher-order layering transitions was not possible within the
accuracy of our numerical procedure.

\begin{figure}
\epsfig{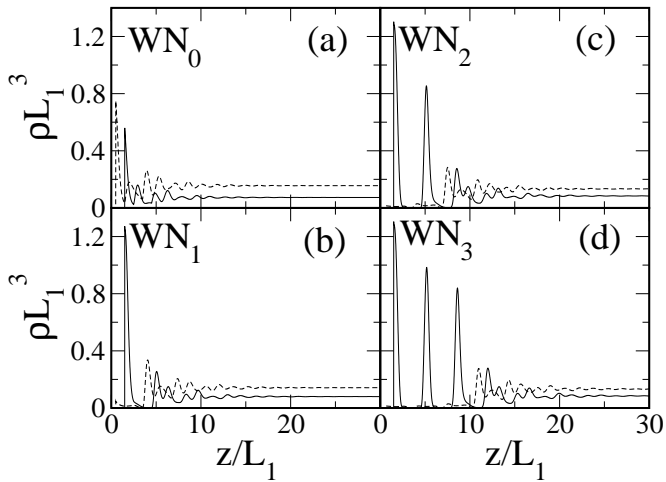}
\caption{Density profiles of species 1 (dashed line) and 2 (solid line) for 
$\beta pL_1^3=1.30$ and (a) $x=0.6800$, (b) $0.6400$, (c) $0.61400$ 
and (d) $0.6120$ (coexistence value of composition is $x_{\rm coex}=0.6115$). 
The symbol WN$_i$ ($i=1,2,3...$) denotes the interfacial structure 
containing $i$ adsorbed layers.}
\label{fig3}
\end{figure}  

In Fig. \ref{fig4} the behavior of the WN surface tension
$\gamma=\gamma_{\hbox{\tiny WN}}$,
as a function of composition, is shown. The location of the WN$_{i-1}$--WN$_i$ 
layering transition is obtained from the intersection of the surface tensions 
corresponding to the two structures. The surface tension of the
WN$_0$ structure, shown in the inset, is somewhat peculiar: just before
the WN$_0$--WN$_1$ layering transition, the surface tension exhibits a maximum.
This behavior can be explained by resorting to Eqn. (\ref{relation}) 
and noting that the surface tension slope is exactly zero 
at $x^*=\Gamma_1/(\Gamma_1+\Gamma_2)
=\overline{\rho}_1/(\overline{\rho}_1+\overline{\rho}_2)$, where 
$\overline{\rho}_i=H^{-1}\int_0^H\rho_i(z)dz$, i.e. when the composition 
of the mixture at bulk coincides with its interfacial value. 
If $x>x^*$, i.e. when the interfacial composition is lower than the bulk 
value, the slope of the surface tension is positive, while the opposite 
occurs for $x<x^*$. 

The adsorption coefficients $\Gamma_i$, defined in (\ref{Ads}),
are plotted in Fig. \ref{fig5} as a function of $x$. As can be seen, $\Gamma_2$ abruptly increases at the layering 
transitions while $\Gamma_1$ abruptly decreases, i.e. the WN interface  
exhibits adsorption of the long species and desorption of the small species. 
This is the natural interfacial path that connects a nematic phase located 
far from the wall, and rich in short particles, with a smectic film located 
next to the wall, and rich in long species, as $x\to x_{\rm{coex}}$.

\begin{figure}
\epsfig{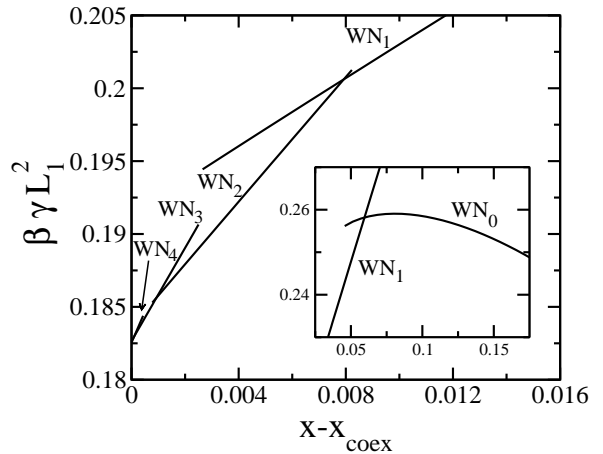}
\caption{Surface tension of the wall-nematic interface 
versus composition $x-x_{\rm{coex}}$ for reduced pressure $\beta pL_1^3=1.3$. 
Symbols WN$_i$ denote the different 
branches corresponding to wall-nematic interfaces containing $i$ adsorbed 
smectic layers. The inset shows the surface tensions for the
WN$_0$-WN$_1$ surface phase transition.}
\label{fig4}
\end{figure}

Repeating the same calculations, but at a lower value of pressure,
$\beta pL_1^3=1.15$, we find that the first two layering transitions 
disappear, while the higher-order transitions WN$_{i-1}$--WN$_i$, with $i>2$,
remain. Fig. \ref{fig6}(b) 
shows that, although the WN$_0$--WN$_1$ and WN$_1$--WN$_2$ 
transitions are absent, the adsorption coefficients significantly increase 
in the neighborhood of the transition points corresponding to a higher 
pressure. This behavior is consistent with the occurrence of critical
points for the WN$_0$--WN$_1$ and WN$_1$--WN$_2$ transitions at critical
pressures in the interval $1.15<\beta pL_1^3<1.30$. 
Fig. \ref{fig6}(a) shows a partial sequence of layering transitions
involving up to 12 surface layers (the maximum number that our
numerical scheme can deal with). It is reasonable to think that the
layering transitions will continue up to the bulk transition in 
infinite number (complete wetting scenario). 

A surface 
phase diagram that includes the first four layering transitions is shown in  
Fig. \ref{fig7}. The following trends can be extracted from the figure:
(i) all layering transition curves approach the nematic binodal as
the pressure is increased (for a number of layers $>5$ the curves are
too close to the nematic binodal and are not visible in the figure);
(ii) the critical points, where layering transitions terminate, 
move to lower pressures as the number of layers increases for $i\ge 2$. 
It is interesting to note that the critical point of the WN$_0$--WN$_1$
transition is located below that of the WN$_1$--WN$_2$ transition; this
feature is related to the strong ordering of the WN interface 
just before the WN$_0$--WN$_1$ transition. In any case, layering
transitions terminate at pressures where the bulk NS demixing transition
becomes weak or disappears, i.e. when $\beta pL_1^2\simeq 1$.

We note that, depending on the wetting scenario for the WS$_2^{\prime}$
interface, the layering transition curves could or could not continue 
above the NS$_2^{\prime}$ spinodal; for example, the wetting r\'egime 
could change to a partial wetting r\'egime, similar to that
found in Ref. \cite{Somoza}. Since interfacial calculations with
a bulk smectic phase are difficult, we have not carried out this
programme in the present work. 

\begin{figure}
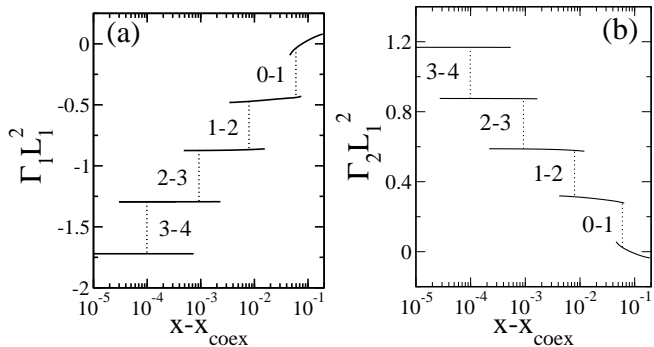

\epsfig{file=Fig5a.eps,width=1.65in}
\epsfig{file=Fig5b.eps,width=1.65in}
\caption{Adsorption coefficients of (a) species 1, (b) species 2, 
as a function of composition $x-x_{\rm{coex}}$ (in logarithmic scale). Dotted lines
indicate the transition points between two equilibrium interfaces.
The value of the reduced pressure is $\beta pL_1^3=1.3$. Labels indicate 
the number of layers of the structures involved in each layering
transition.}
\label{fig5}
\end{figure}  

\begin{figure}
\epsfig{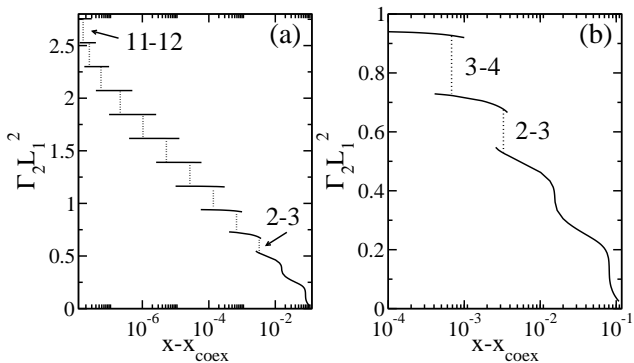}
\caption{(a) Adsorption coefficient of species 2 as a function of 
$x-x_{\rm{coex}}$ (in logarithmic scale). Equilibrium structures are
represented by continuous curves, while dashed vertical lines
indicate layering transitions (metastability branches
are not meant to be complete). (b) A zoom showing the first two layering
transitions (in this case adsorption of the first two layers does not proceed 
via surface phase transitions). The pressure is fixed at $\beta pL_1^3=1.15$.
Labels indicate the number of layers of the structures involved in each layering
transition.}
\label{fig6}
\end{figure}

Next we briefly discuss the transition strength along
the layering transition curves. In Fig. \ref{fig8} the gap in the adsorption 
coefficient of species 2 at coexistence of the WN$_{i-1}$ and WN$_{i}$ 
structures, $\Delta\Gamma_2^{(i)}
=\Gamma_2^{(\hbox{\tiny WN}_{i})}-\Gamma_2^{(\hbox{\tiny WN}_{i-1})}$, is plotted
as a function of composition along the layering 
transition curves and for various indices $i$. 
The general trend observed is that, as more layers get 
involved, the transition becomes stronger (i.e. the 
gap at coexistence is larger). As the index $i$ of the layering 
transition increases, the gap $\Delta\Gamma_2^{(i)}$ tends to saturate,
corresponding to the fact that the additional layers adsorbed do not
feel the effect of the wall and therefore contribute to the adsorption
coefficient with a constant quantity.
  
\begin{figure}
\epsfig{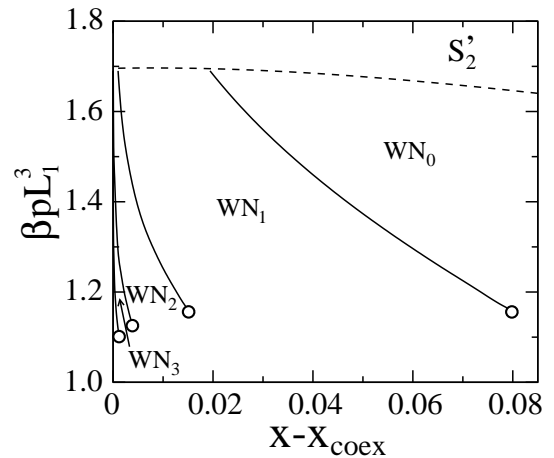}
\caption{Layering transitions (solid curves) 
between WN$_{i-1}$ and WN$_i$ interfacial structures (with $i$ the number of 
adsorbed layers) in the reduced pressure--composition plane. 
The critical points of the transitions 
are shown with open circles. Dashed line is the NS$_2^{\prime}$
second order transition.}
\label{fig7}
\end{figure}

\begin{figure}
\epsfig{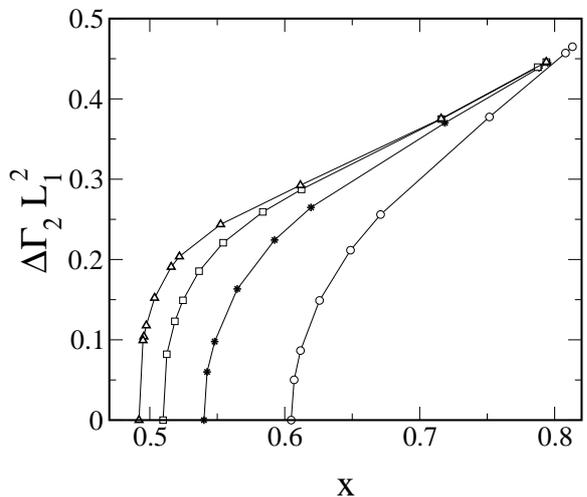}
\caption{Coexistence gap of the adsorption coefficient of the 
second species, $\Delta\Gamma_2^{(i)}$, at the WN$_{i-1}$--WN$_{i}$ 
layering transitions as a function of composition and for various
indices $i$. Symbols refer to layering transitions WN$_0$--WN$_1$ (circles),
WN$_1$--WN$_2$ (stars), WN$_2$--WN$_3$ (squares) and
WN$_3$--WN$_4$ (triangles).}
\label{fig8}
\end{figure}

We end this section with a comment on the origin of the layering 
transitions. As shown in \cite{M-R3}, two-dimensional 
one-component hard-rod fluids in contact with a hard wall do not exhibit 
layering transitions even though the bulk transition is of first order \cite{M-R2}. 
This is also probably the case in the corresponding three-dimensional fluid,
although we have not performed explicit calculations for the current model.
Therefore, one tentative explanation for the phenomenology found in the mixture 
is that layering transitions are due to the entropic coupling between the two 
species mediated by the hard wall: next to the wall, where particle densities are 
high, both species compete for the gain in entropic volume. Just above the layering
transition, a mixed layer packs less efficiently, and short particles are 
abruptly depleted from that region, with the subsequent abrupt increase in 
long particles. This conclusion would not be affected by the discovery of layering
transitions in the calculations of Somoza et al. \cite{Somoza}, who use
particles with additional soft, temperature-driven interactions; in this fluid
the mechanism behind the layering transitions would be completely different. 

A look at the structure of the density profiles of the
WN$_0$ interface [Fig. \ref{fig3}(a)] can help us understand this effect
from a different perspective.
The density maxima of the two species are clearly separated, due to the
different lengths of the particles (density is maximum exactly at contact
with the wall). However, the maxima of the bulk smectic phase are in phase
(S$_1$ smectic). It is only because a sharp change in the interfacial
structure occurs, via a first-order layering transition, that the
interface can relax to a structure compatible with that in the bulk, i.e.
with the correct relative phase. This mechanism operates even for structures 
WN$_i$ with large $i$, when the effect of the wall is not crucial, 
because the density maxima of the two species, in the region between
the already-formed 
smectic layers and the nematic, are always displaced one with respect
to the other. At lower pressures, such that the bulk smectic phase is
S$_2$, there is no such incompatibility between the bulk structure and
the structure imposed by the wall, and the layering transitions vanish.

\subsection{Wetting behavior}
\label{wetting}

To obtain a global picture of the wetting behavior of the mixture, we have 
concentrated on four different values of reduced pressure: $\beta pL_1^3=1.25$
and $1.30$ (located above the bulk triple point, see Fig. \ref{fig1}), $1.00$ 
(below the triple point and above the tricritical point), and $0.495$ 
(below the tricritical point). In the first two cases we have found the 
phenomenology described in Sec. \ref{layering}, i.e. an infinite sequence of 
layering transitions leading to complete wetting of the WN interface by the S$_1$ phase. 
Using the procedures described in Sec. \ref{w-f}, we have calculated
the surface tensions of the WN, WS$_1$ and NS$_1$ interfaces for
$x=x_{\rm{coex}}$, which are necessary to
discuss the wetting behaviour. Their values are collected in Table \ref{tab1}. 
As corresponds to complete wetting by the S$_1$ phase, the surface
tensions fulfill Young's law 
$\gamma_{\hbox{\tiny WS$_1$}}=\gamma_{\hbox{\tiny WN}}+
\gamma_{\hbox{\tiny NS$_1$}}$ (the value of $\gamma_{\hbox{\tiny WN}}$ 
at $x=x_{\rm{coex}}$ can be computed by extrapolation 
of $\gamma_{\hbox{\tiny WN}_i}$ with $i\to\infty$. In
practice $i=32$ already gives enough accuracy to assess the wetting
behaviour. Note that, in these cases, all the WN$_i$ structures are 
metastable and can be stabilised, even at coexistence, under conditions
of complete wetting, i.e. when the absolute free-energy minimum actually 
corresponds to $i=\infty$).

\begin{table}
\begin{tabular}{||cccccc||}
\hline
\hline
$\beta pL_1^3$ & $x_{\rm coex}$ & $\gamma_{\rm{WS_j}}^*$ & 
$\gamma_{\rm{NS_j}}^*$ & 
$\gamma_{\rm{WN}}^*$ & $\gamma_{\rm{WS_j}}^*+\gamma_{\rm{NS_j}}^*$\\
\hline
1.300 & 0.61150 & 0.121166 & 0.061370 & 0.182535 & 0.182536\\
1.250 & 0.58231 & 0.122408 & 0.055451 & 0.177859 & 0.177859\\
1.000 & 0.43815 & 0.155570 & 0.001243 & 0.156822 & 0.156813\\
0.495 & 0.10000$^{*}$ & 0.181061& -- & 0.181061 & --\\
\hline
\hline
\end{tabular}
\caption{Reduced surface tensions $\gamma^*=\beta\gamma L_1^2$ 
of the WS$_j$, WN and NS$_j$ interfaces for different values of the reduced 
pressure. Here $j=1,2$ depending on the nature of the smectic phase.
$^*$ indicates value at spinodal.}
\label{tab1}
\end{table}

The wetting behavior for $\beta pL_1^3=1.00$ is similar to that found 
in \cite{Dani2} and \cite{M-R3} for one-component hard-rod systems: 
the thickness of the smectic film adsorbed at the WN interface grows 
continuously as $x\to x_{\rm{coex}}$ and diverges at the bulk transition.
This behavior is illustrated in Fig. \ref{fig9}, where four density
profiles for values of composition very close to the bulk transition
are shown. In this case layering transitions are completely absent;
adsorption coefficients $\Gamma_i$ as a function of $x$ do not exhibit
any discontinuity (Fig. \ref{fig10}), but diverge logarithmically as
$x\to x_{\rm{coex}}$. Young's law for complete wetting is also fulfilled 
within the numerical accuracy that could be achieved in this case 
(see Table \ref{tab1}). In this case the surface tension 
$\gamma_{\rm{NS_1}}$ is very small and is subject to higher uncertainties
(the value of $\gamma_{\rm{WN}}$ was obtained by extrapolation to
coexistence, $x\to x_{\rm coex}$).

\begin{figure}
\epsfig{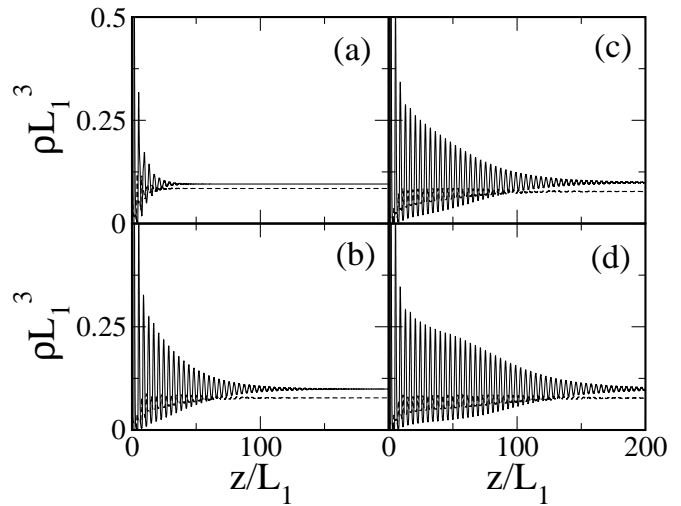}
\caption{Density profiles of species 1 (solid line) and 2 (dashed line) 
of the WN interface at bulk pressure $\beta pL_1^3=1.00$ and values
of composition (a)
$\Delta x=x-x_{\rm{coex}}=3.2\times 10^{-2}$, (b) $1.3\times 10^{-3}$,
(c) $3.2\times 10^{-4}$ and (d) $6.4\times 10^{-5}$. The first and,
except in (a), the second density peaks are truncated due to the
small scale of the vertical axis.}
\label{fig9}
\end{figure}

\begin{figure}
\epsfig{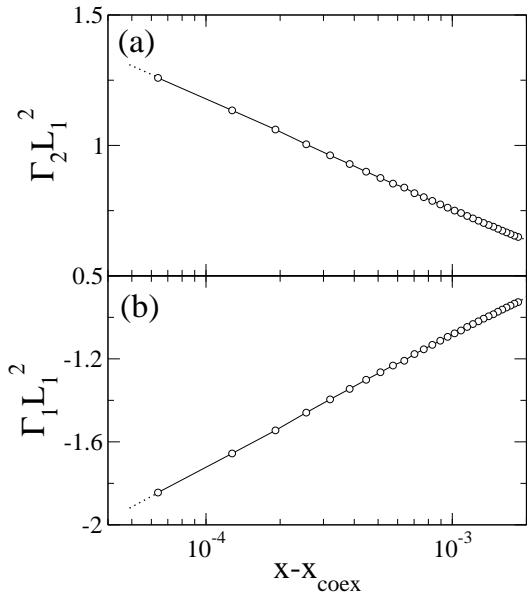}
\caption{ Adsorption coefficients of species 2 (a) and 1 (b) as a 
function of composition (in logarithmic scale) near the wetting transition. 
The reduced bulk pressure is $\beta pL_1^3=1.0$.}
\label{fig10}
\end{figure}

Next we discuss the equilibrium density profiles of 
the WN interface at a pressure $\beta pL_1^3=0.495$ (i.e. below the 
tricritical point) and, more specifically, 
the behavior of the adsorption coefficients as the 
bulk NS spinodal is approached. Let $x^*$ be the composition of the
spinodal at a given pressure. Since the NS transition is of second order, 
we should find critical adsorption, similar to that 
occurring at the liquid-vapour critical point where the adsorption
diverges logarithmically as dictated by mean-field theory. 
Far from the wall and close to the bulk spinodal, the WN interface 
exhibits oscillations with a period $d^*$ (the smectic period at bifurcation). 
Thus, the deviation of density profiles from their bulk values is better 
accounted for by the quantity $|\rho_i(z)-\rho_i|$ (the analogue of the 
order parameter in the anti-ferromagnetic Ising model), and it is
convenient to define modified adsorption coefficients as
\begin{eqnarray}
\Gamma_i^*=\int_0^H|\rho_i(z)-\rho_i|dz.
\end{eqnarray}
The behaviors of $\Gamma_2$ and $\Gamma_2^*$ as a function of $x-x^*$  
are illustrated in Fig. \ref{fig11}. It can be seen from the figure 
that, while $\Gamma_2$ seems to reach a plateau as $x\to x^*$, 
the modified coefficient $\Gamma_2^*$ diverges logarithmically as 
predicted by mean-field theory for a critical adsorption phenomenon \cite{comment}. 
The value of the plateau is difficult to determine due to the huge values of $H$
required to accommodate the weakly damped interfacial oscillations that
extend very far from the wall when $x\sim x^*$. The range of these oscillations
is of the order of the smectic bulk correlation length, which diverges at 
$x=x^*$. Finally, we have checked that the system also exhibits 
critical adsorption in the neighbourhood of the NS$^{\prime}$
spinodal curve (at higher pressure and composition).

\begin{figure}
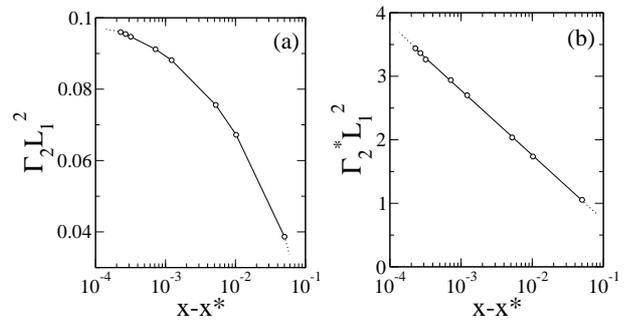

\epsfig{file=Fig11a.eps,width=1.6in}
\epsfig{file=Fig11b.eps,width=1.5in}
\caption{The original (a) and modified (b) adsorption coefficients of 
second species as a function of bulk composition near the NS second order  
transition at $\beta pL_1^3=0.495$. In (a) the curve is a guide to the eye.
In (b) the line is a {logarithmic fit.}}
\label{fig11}
\end{figure}

We end this section with a discussion on the impact of the parallel-particle
approximation on the wetting behaviour. Complete wetting (either continuous 
or via a sequence of layering transitions) of a hard wall by a binary mixture 
of hard particles is governed by two factors: (i) the effective entropic
interactions between particles and the wall, and (ii) the distance of the bulk
state point from the demixing binodal. Since both the free- and
restricted-orientations models contain these two features, we only expect quantitative
deviations between the two as far as the wetting behaviour is concerned.
The situation with respect to the critical adsorption phenomenon is different, because 
this is due to the second-order character of the nematic-smectic transition, which may 
become weakly first order for small perturbations of particle orientations with respect
to perfect alignment. In a model with free orientations the critical adsorption
behaviour could be superseded by complete (continuous) wetting. 

\section{Conclusions}
\label{conclusions}

In this work we have studied the surface adsorption phenomena of a  
liquid-crystal colloidal mixture that has a stable smectic phase at 
moderate pressures. The mixture is described by means of a very simple model
consisting of perfectly aligned hard particles, while the substrate 
is a hard wall inducing perfect homeotropic anchoring and nematic boundary
conditions far from the wall are chosen. Even with these simple
assumptions, the bulk and surface phase diagrams are so rich that
we have concentrated only on the analysis of a single mixture with 
length ratio $s=L_2/L_1=3$. The theoretical tool used is a recently 
developed fundamental-measure density functional for mixtures of 
parallel cylinders \cite{M-R1}.

We have found a bulk phase diagram with second order NS transitions 
at low pressures, followed by NS demixing above a tricritical point. 
In the low and high composition regions of the phase diagram two
critical points exist, above which two smectic phases, one of them
micro-fractionated, coexist. Coexistence is ended by corresponding
triple points at higher pressures. At the highest pressures investigated
smectic demixing is found, with each smectic rich in 
one of the species. A bifurcation analysis corroborates that the NC 
spinodal is always above the NS spinodal, but does not completely
clarify the question about the absolute stability of the smectic 
against the columnar phases. In any case, we do not expect the 
NS, and possibly also the SS, demixing transitions to be preempted by the 
columnar phase at low pressures.
 
The surface phase diagram has three different wetting r\'egimes. 
The first one, located below the tricritical point, exhibits critical 
adsorption as the composition of the bulk nematic phase approaches the
NS spinodal. In the second r\'egime, located approximately 
above the tricritical point and below the triple point 
(the exact boundaries would require further analysis), there exists 
complete wetting of the
substrate by a smectic film whose thickness diverges logarithmically as 
$x\to x_{\rm{coex}}$. Finally, the third r\'egime is located above the 
triple point and is characterized by the presence of layering transitions
that ultimately lead to complete wetting. A previous
theoretical study of one-component hard-rod fluids using a density 
functional model
\cite{Somoza} found layering phenomena in the semi-infinite system due 
to strong attractive interactions between the wall and the fluid particles. 
By contrast, layering transitions in hard-rod liquid-crystal mixtures adsorbed on 
a hard wall, as shown in the present study, is a direct consequence of the 
wall-mediated entropic interaction between the two species. 

We expect that the present work serves as a starting point to initiate 
experimental studies on the surface phase behavior of liquid-crystal 
colloidal mixtures consisting of particles that interact through 
short-ranged repulsive forces, and having a stable bulk smectic phase. 
These experiments could be guided by the phenomenology
found in the present study. Our future theoretical studies will analyse
the adsorption phenomenology of films in the neighbourhood of
the bulk triple point, a challenging problem that could provide 
further interesting phenomena.

\section*{Acknowledgments}
We gratefully acknowledge financial support from Comunidad 
Aut\'onoma de Madrid (Spain)
under grants NANOFLUID, MOSAICO and 
S-0505/ESP-0299. This work is partly financed by grants 
FIS2005-05243-C02-01, FIS2007-65869-C03-01, FIS2008-05865-C02-02 
and FIS2007-65869-C03-C01 from Ministerio de Educaci\'on y 
Ciencia (SPAIN).

\appendix

\section{Bifurcation analysis}
\label{bifurca}
To implement the bifurcation analysis, we use the following expressions for the 
Fourier transforms of the direct correlation functions \cite{M-R1,M-R4}:
\begin{eqnarray}
&&\rho_i\hat{c}_{ij}({\bf q},\eta)=x_i\left\{2l_{ij}\Psi_0(q_{\perp})
\chi\left(q_{\parallel} l_{ij}\right)\right.\nonumber\\\nonumber\\
&&\left.+l_il_j\Psi_1(q_{\perp})
\chi\left(q_{\parallel}l_i/2\right)
\chi\left(q_{\parallel}l_j/2\right)\right\},
\end{eqnarray}
where $l_i=L_i/\langle L\rangle$, $l_{ij}=(l_i+l_j)/2$, $\langle L\rangle=\sum_i x_iL_i$,
with $q_{\parallel}$ and $q_{\perp}$ in units of $\langle L\rangle$ and $D/2$, 
respectively. We have defined $\chi(x)\equiv\sin (x)/x$ and  
\begin{eqnarray}
\Psi_0(q)&=&4y\left\{\frac{J_1(2q)}{q}+2yJ_0(q)\frac{J_1(q)}{q}
\right.\nonumber\\
&&\left.+y(1+2y)\left[\frac{J_1(q)}{q}\right]^2\right\},\\  
\Psi_1(q)&=&4y^2\left\{\frac{J_1(2q)}{q}+2(1+2y)J_0(q)
\frac{J_1(q)}{q}\right.\nonumber\\
&&\left.+(1+6y+6y^2)\left[\frac{J_1(q)}{q}\right]^2
\right\},
\end{eqnarray}
with $y=\eta/(1-\eta)$ and $J_n(x)$ the n-th order Bessel function of first kind. 
The NS spinodal can be obtained by solving Eqs. (\ref{set}) with
${\bf q}_{\perp}={\bf 0}$, $q_{\parallel}=q$, and
\begin{eqnarray}
&&{\cal H}(q,\eta)=1+\sum_i x_i\left\{2\Psi_0(0)l_i\chi(ql_i)
\right.\nonumber\\&&+
\left.\Psi_1(0)l_i^2\chi^2(ql_i/2)\right\}-\left[\Delta\Psi_0(0)\chi(q\Delta l/2)
\right]^2,
\label{spino}
\end{eqnarray}
where $\Delta l=l_1-l_2$, $\Delta^2=\langle l^2\rangle-1$ being the
the polydispersity coefficient, with $\langle l^2\rangle=\sum_ix_il_i^2$.
For columnar symmetry $q_{\parallel}=0$, $|{\bf q}_{\perp}|=q$, and we find
\begin{eqnarray}
{\cal H}(q,\eta)=1+2\Psi_0(q)+\Psi_1(q)+\Delta^2\left[\Psi_1(q)
-\Psi_0^2(q)\right].\nonumber\\
\end{eqnarray}
To search for a possible NN demixing scenarios, we need to solve ${\cal H}(0,\eta)=0$ 
for $\Delta$ as a function of $\eta$, which provides the NN spinodal: 
\begin{eqnarray}
\Delta^2=\frac{1+2\Psi_0(0)+\Psi_1(0)}{\Psi_0^2(0)-\Psi_1(0)}
=\left(\frac{1}{\eta}-1\right)^2\frac{1+4\eta+\eta^2}{
7-2\eta-\eta^2}.
\label{ee} \nonumber\\
\end{eqnarray} 
For our particular mixture ($L_2=3L_1$), the pressure of the NN critical point is 
$\beta p L_1^3\approx 10$. This result shows the metastable character of the NN 
demixing against NS demixing (see Sec. \ref{bulk}).

In Fig. \ref{figA1} the NS and NC spinodals, in the pressure--composition plane,
are shown. They never intersect and the former is always below the latter.
{This fact indicates that in the low-pressure region of the phase diagram, 
calculated in Sec. \ref{bulk}, the nematic and smectic phases could 
actually be stable against the columnar phase.}

\begin{figure}
\epsfig{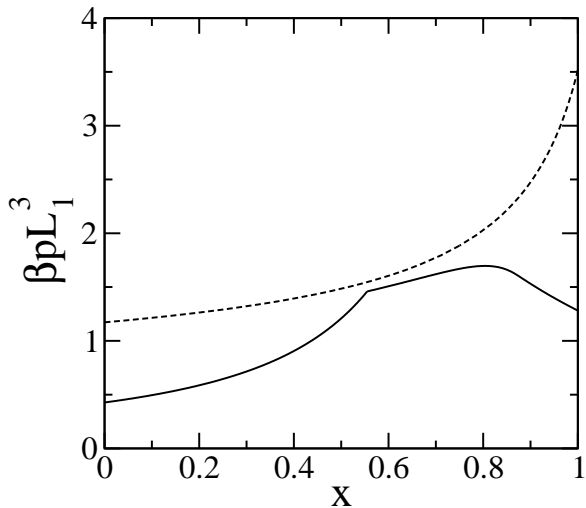}
\caption{NS (solid) and NC (dashed) spinodals in the reduced pressure--composition 
plane.}
\label{figA1}
\end{figure}

\section{Interfacial Gibbs-Duhem relation}
\label{G-D}
In this section we obtain the thermodynamic relation (\ref{relation}) involving the
derivative of the WN surface tension with respect to composition and
the adsorption coefficients $\Gamma_i$. Starting from the interfacial Gibbs-Duhem 
relation $d\gamma=-\sum_i\Gamma_id\mu_i$, we have
\begin{eqnarray}
\frac{d\gamma}{dx}=-\sum_i\Gamma_i\frac{d\mu_i}{dx}.
\label{GD}
\end{eqnarray}
For fixed pressure $p_0$ the chemical potentials $\mu_i(x,\rho(x))$ are only functions of 
$x$ because $\rho(x)$ is defined implicitly through  
the constraint $p(x,\rho(x))=p_0$. Then:
\begin{eqnarray}
\frac{d\rho}{dx}=-\frac{\partial p/\partial x}{\partial p/\partial \rho}.  
\label{jj}
\end{eqnarray}
Taking into account that
\begin{eqnarray}
\frac{d\mu_i}{dx}=\frac{\partial \mu_i}{\partial x}+\frac{\partial \mu_i}
{\partial\rho}\frac{d\rho}{dx},
\end{eqnarray}
and using Eqs. (\ref{chepo}), (\ref{press}) and (\ref{jj}), we finally find 
\begin{eqnarray}
\beta\frac{d\mu_i}{dx}=\frac{(-1)^{i-1}}{x_i}
-\frac{\Delta l\eta^2(7-2\eta-\eta^2)(1-l_i)}{
(1+4\eta+\eta^2)(1-\eta)^2}
\end{eqnarray}
which, after insertion in (\ref{GD}), gives the final result (\ref{relation}), with 
\begin{eqnarray}
{\cal U}(x,p)=1-\frac{\eta^2(7-2\eta-\eta^2)\Delta^2}
{(1+4\eta+\eta^2)(1-\eta)^2}.
\label{oo}
\end{eqnarray}
The function $\eta(x)$ in Eq. (\ref{oo}) can be found 
from the constant-pressure constraint. Comparing Eqs. (\ref{ee}) and (\ref{oo}), 
which contain the same and only density factor that can change sign, we 
conclude that ${\cal U}(x,P)\geq 0$ only when the mixture is stable against NN 
bulk demixing. To check for consistency of our numerical minimization procedure, 
we compare $d\gamma/dx$, as calculated from Eq. (\ref{relation}) [i.e. using the  
adsorption coefficients $\Gamma_i$ obtained from the equilibrium density profiles
$\rho_i^{(\rm{e})}$] with the numerical derivative with respect to $x$ of the 
surface tension $\gamma$ obtained after minimization. Both results are plotted 
in Fig. \ref{check} (for the WN interface) for $\beta pL_1^3=1.3$. As can be seen,
both methods reproduce the same function with high accuracy, which demonstrates
that our calculations are fully consistent. Note that the slope of the  
surface tension is equal to zero (and consequently $\gamma$ has a maximum as 
a function of $x$) for  $x=\Gamma_1/(\Gamma_1+\Gamma_2)$ [see Eq. (\ref{relation})],
i.e. when the bulk composition is identical to the relative fraction of adsorption 
coefficient of species 1.

\begin{figure}
\vspace*{0.7cm}
\epsfig{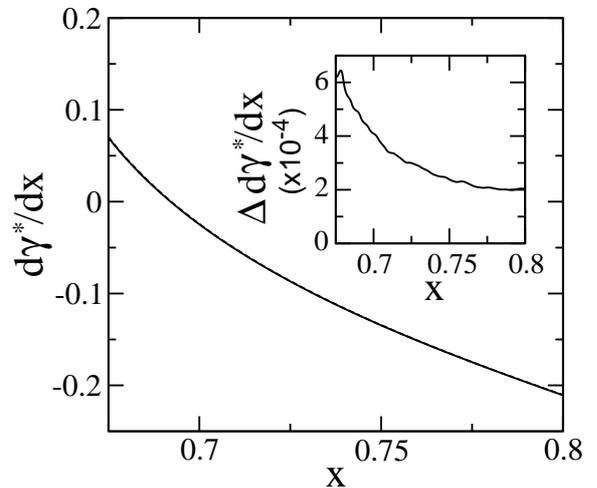}
\caption{Derivative of the reduced surface tension of the WN interface, 
$\gamma^*=\beta\gamma_{\hbox{\tiny WN}}L_1^2$, 
with respect to composition $x$, obtained from numerical differentiation
of the function $\gamma(x)$. In the inset the difference between this derivative
and that obtained from Eqn. (\ref{relation}) is plotted (wiggles are due to 
noise in the numerical derivative). The value of the reduced pressure is 
$\beta pL_1^3=1.3$.}
\label{check}
\end{figure}


\begin{thebibliography}{13}
\bibitem{comment} The bifurcation point for $\beta pL_1^3=0.495$ is at 
$x^*=0.10000$ as obtained numerically from Eqns. (\ref{set}) and (\ref{spino}).
The data for $\Gamma_2^*$ in Fig. \ref{fig10} are very sensitive to 
$x-x^*$ close to the bifurcation point, especially when plotted in logarithmic
scale. An unwanted consequence of this is that the first two values in
Fig. \ref{fig10} exhibit a slight curvature if the above value of $x^*$ is used.
This situation is improved if $x^*$ is obtained from a fit to the function
$\Gamma_2^*=-a\log{(x-x^*)}-b$, with $a,b$ and $x^*$ free parameters.
The fit gives $x^*=0.09987$. The data plotted in
the figure were obtained using this modified value.
\end{thebibliography}
\end{document}